\documentclass[fleqn,twoside]{article}
\usepackage{espcrc2}

\usepackage{graphicx}
\usepackage[figuresright]{rotating}

\newcommand{\AmS}{{\protect\the\textfont2
  A\kern-.1667em\lower.5ex\hbox{M}\kern-.125emS}}
\def\Journal#1#2#3#4{{#1} {\bf #2}, #3 (#4)}

\def\NIMA{{Nucl. Instrum. Methods} A}

\def\PLB{{Phys.Lett.}B}
\def\PRL{Phys. Rev. Lett.}
\def\PRD{{Phys. Rev.} D}

\def\APJ{Astrophys. J.}
\def\SNP{Sov. Jour. Nucl. Phys.}
\newcommand{\plumin}[2]{^{+#1}_{-#2}}

\def\dmsq{\Delta m^2}
\def\sisq{\sin^2\theta}
\def\tasq{\tan^2\theta}

\title{The Solar Neutrino Day/Night Effect in Super-Kamiokande}

\author{Michael B. Smy {\em for the Super-Kamiokande Collaboration}\\
Department of Physics and Astronomy, University of California, Irvine,
CA 92697-4575}
       
\begin{document}

\begin{abstract}
The time variation of the elastic scattering rate of
solar neutrinos with electrons in Super-Kamiokande-I was fit
to the day/night variations expected from active two-neutrino
oscillations in the Large Mixing Angle region.
Combining Super-Kamiokande measurements with other solar and reactor neutrino data,
the mixing angle is determined as $\sisq=0.276\plumin{0.033}{0.026}$
and the mass squared difference between the two neutrino mass
eigenstates as
$\dmsq=7.1\plumin{0.6}{0.5}\times10^{-5}$eV$^2$.
For the best fit parameters, a day/night asymmetry
of $-1.7\pm1.6$(stat)$\plumin{1.3}{1.2}$(syst)\%
was determined from the Super-Kamiokande data, which
has improved statistical precision over previous measurements
and is in excellent agreement with the expected value
of $-1.6\%$.
\vspace{1pc}
\end{abstract}

\maketitle

\section{Introduction}

The combined analysis of all solar neutrino experiments
\cite{solar} gives firm evidence for neutrino oscillations.
All data are well described using just two neutrino mass eigenstates
and imply a mass squared difference between $\dmsq=3\times10^{-5}$eV$^2$
and $\dmsq=1.9\times10^{-4}$eV$^2$ and a mixing angle between 
$\tasq=0.25$ and $\tasq=0.65$~\cite{global}. This region
of parameter space is referred to as the Large Mixing Angle solution
(LMA). The rate and spectrum of reactor anti-neutrino interactions
in the KamLAND experiment~\cite{kl} are also well reproduced for these
mixing angles and some of these $\dmsq$.
Over the $\dmsq$ range of the LMA, solar $^8$B neutrinos are
$\approx$100\% resonantly converted into the second mass eigenstate
by the large matter density inside the sun~\cite{MSW}.
Therefore, the survival probability into $\nu_e$ is
$\approx\sin^2\theta$. However, due to the presence of the earth's matter
density, the oscillation probability at an experimental site on earth
into $\nu_e$ differs from $\sin^2\theta$ during the night. Since
Super-Kamiokande experiment is primarily sensitive to $\nu_e$'s, this induces
an apparent dependence of the measured neutrino interaction rate
on the solar zenith angle (often a regeneration of $\nu_e$'s during the night).
Recently, Super-Kamiokande employed
a maximum likelihood fit to the expected solar zenith angle
dependence on the neutrino interaction rate~\cite{skdn}.
Herein, the statistical uncertainty
was reduced by 25\% compared to previous measurement of the day/night
asymmetry~\cite{global} which consists of two flux measurements in
two separate data samples (day and night). It would require almost three
more years of running time to obtain a similar uncertainty reduction.
Also the GNO, SAGE, and SNO collaborations~\cite{solar}
reported updated neutrino interaction rates.

Super-Kamiokande (SK) is a 50,000 ton water Cherenkov detector described
in detail elsewhere~\cite{skdet}.
SK measures the energy, direction, and time of the
recoil electron from elastic scattering of solar neutrinos with electrons
by detection of the emitted Cherenkov light.
Super-Kamiokande started taking data in April, 1996.
In this report, the full SK-I low energy data set consisting of
1496 live days
(May $31^{\mbox{st}}$, 1996 through July  $15^{\mbox{th}}$, 2001)
is used.

\begin{figure}
\includegraphics[width=3in]{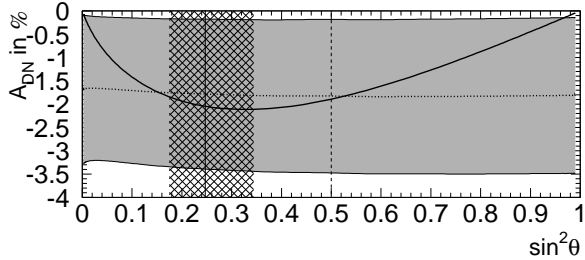}
\caption{Fitted SK Day/Night Asymmetry as a Function of Mixing.
The $\dmsq$ is $6.3\times10^{-5}$eV$^2$.
The gray band is the $\pm1\sigma$ SK measurement.
The hatched area corresponds to the $\pm1\sigma$ uncertainty of
the $^8$B flux by Junghans et al~\cite{BP2000}. The solid black line shows
the oscillation prediction of the day/night asymmetry.}
\label{fig:dnsin}
\end{figure}

\section{Day/Night Asymmetry}
\begin{figure}
\includegraphics[width=3in]{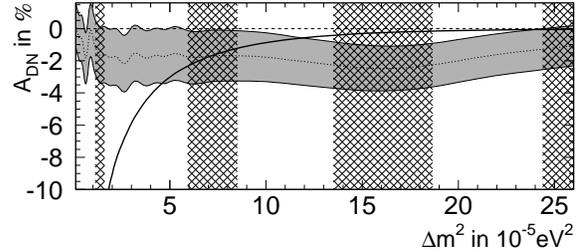}
\caption{Fitted SK Day/Night Asymmetry as a Function of $\dmsq$.
The gray band is the $\pm1\sigma$ SK measurement.
The hatched area corresponds to the 95\% allowed contours reported
by the KamLAND collaboration. The solid black line shows
the oscillation prediction of the day/night asymmetry.}
\label{fig:dndm}
\end{figure}

The solar zenith angle $\theta_z$ between the solar direction
and the vertical direction defines the path length of the solar neutrino
inside the earth. During the day ($\cos\theta_z<0$) this path length
is zero, during the night ($\cos\theta_z>0$) it varies between zero
and (up to) the diameter of the earth.
The day/night rate asymmetry is defined as
\[
A_{\mbox{\tiny DN}}=\frac{D-N}{0.5(D+N)}
\]
where $D$ ($N$) refers to the average neutrino interaction rate during
the day (night). If the neutrino interaction rate during the night
varies significantly from the average night rate $N$, and if the
functional form (shape) of this variation is known, the amplitude of
this time variation of the rate can be determined more accurately
then just calculating $A_{\mbox{\tiny DN}}$
from the average rates. These conditions are met for two-neutrino oscillations
in the LMA region. In~\cite{skdn} a maximum likelihood fit to
the SK data finds a day/night amplitude equivalent to
$A_{\mbox{\small DN}}=-1.8\pm1.6$(stat)$\plumin{1.3}{1.2}$(syst)\%.
The fit assumes $\dmsq=6.3\times10^{-5}$eV$^2$
and $\tasq=0.55$.
The asymmetry calculated from the measured average day and night rates
on the other hand is
$A_{\mbox{\small DN}}=-2.1\pm2.0$(stat)$\plumin{1.3}{1.2}$(syst)\%~\cite{global}.
It assumes a step function for the time variation and therefore does not
reflect any oscillation parameters.
The dependence of the fitted day/night amplitude on the mixing angle
$\sisq$ is shown in Figure~\ref{fig:dnsin}. Overlaid are the predicted asymmetries
and the solar model constraint of the $^8$B neutrino flux from
Junghans et al~\cite{BP2000}.
The $\dmsq$ dependence is stronger
as can be seen in Figure~\ref{fig:dndm}. Overlaid are the
predicted asymmetries and bands (typically called
LMA-0, LMA-I, LMA-II, etc) corresponding to
the KamLAND 95\% allowed contours: the SK day/night
measurement excludes LMA-0, and favors LMA-I.

\section{Full Oscillation Analysis}
\begin{figure}
\noindent\hspace*{-0.18in}\includegraphics[width=3.15in]{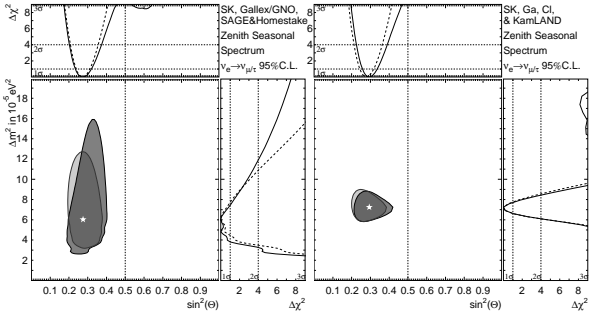}
\caption{Allowed Regions at 95\% C.L. from solar neutrino data
(left) and solar neutrino \& KamLAND (right) reactor neutrino data.
The functions at the top and right of each panel are the mariginalized
$\Delta\chi^2$ distributions. 
The dark gray areas/solid lines
use the solar neutrino data from Gallex/GNO,
SAGE, Homestake \& SK the SSM neutrino fluxes
(and the Junghans $^8$B flux constraint),
the light gray areas/dashed lines) solar measurements from
SK \& SNO and no neutrino flux constraints from solar models.}
\label{fig:sol3}
\end{figure}

An oscillation analysis of the SK data by itself is
found in~\cite{skdn}. It describes the solar zenith angle
variation with a likelihood, while the spectrum is fit
with a $\chi^2$ method.
Since the combined solar neutrino
oscillation analysis of~\cite{skdn} was performed,
the neutrino interaction rate measurements of several experiments
improved in precision. In particular, the SNO collaboration
reported a more precise neutral-current interaction rate on
deuterium employing salt to enhance neutron detection.
Figure~\ref{fig:sol3} shows in (dark gray) 
the allowed regions at 95\%
C.L. resulting from the combination of experimental data
from Gallex/GNO, SAGE, the Homestake experiment and SK.
It relies on the $^8$B flux from Junghans and six
low energy neutrino fluxes of the standard solar
model~\cite{BP2000}. Also shown is a combined fit
to SK data, the new salt-enhanced SNO rate measurements, and
the SNO day/night asymmetry. This fit does not rely
on any neutrino flux prediction. Both analyses
yield a unique allowed region -- the LMA solution --
and agree very closely in mixing. The SK/SNO
analysis provides somewhat stronger constraints on
$\dmsq$.
Assuming CPT invariance, both fits are then combined
with a binned likelihood
analysis~\cite{ianni} of the KamLAND reactor anti-neutrino
measurements~\cite{kl}, the results of which are shown
in the right panel. In either case, only LMA-I remains
allowed.

\begin{figure}
\noindent\hspace*{-0.18in}\includegraphics[width=3.15in]{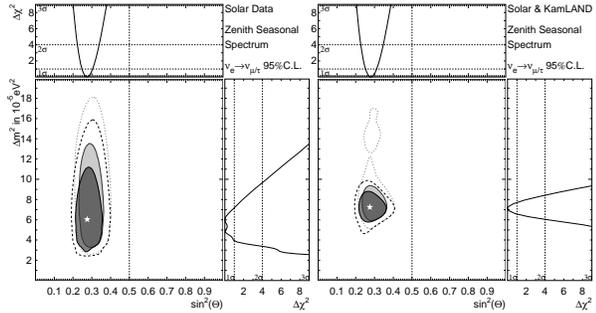}
\caption{Allowed Regions at 95\% (dark gray, solid) and 99.73\% C.L. (dashed)
from solar data (left) and solar \& KamLAND measurements (right).
The solar data includes the SNO salt-phase measurements. Overlaid
are the corresponding regions reported by the SNO collaboration
(light gray and dotted contours) with a weaker $\dmsq$ limit.
The functions at the top and to the right of each panel are the
mariginalized $\Delta\chi^2$ functions.}
\label{fig:osc}
\end{figure}

SNO has also published a combined oscillation analysis,
which uses the SK zenith spectrum $\chi^2$ instead of the
likelihood employed in this report. Figure~\ref{fig:osc}
compares allowed areas of the combined fit to all data 
using the SK likelihood (dark gray areas)
with SNO's contours at 95\% C.L. and $3\sigma$: the $\dmsq$
constraints get stronger when the SK zenith spectrum is replaced
by the SK likelihood. When combined with KamLAND, the
LMA-I is favored over all other solutions by $3.5\sigma$.
The $3\sigma$-allowed LMA-II contour from SNO's analysis
disappears, when the SK likelihood is used.
The oscillation $\chi^2$ is Gaussian;
the parameters are determined as 
$\dmsq=7.1\plumin{0.6}{0.5}\times10^{-5}$eV$^2$ and
$\sisq=0.276\plumin{0.033}{0.026}$. At those
parameters, the day/night asymmetry is expected to
be $-1.6\%$ while the amplitude fit to SK data yields
$-1.7\pm1.6$(stat)$\plumin{1.3}{1.2}$(syst)\%.

\end{document}